\documentclass[aps,prl,amsfonts,amssymb,twocolumn,amsmath,preprintnumbers,floatfix,superscriptaddress]{revtex4-1}%
\usepackage{siunitx}
\usepackage{graphicx}
\usepackage{mathrsfs}
\usepackage{bm}
\usepackage{amsmath}
\usepackage{dcolumn}
\usepackage{epstopdf}
\usepackage{dsfont}
\usepackage{amssymb}
\usepackage{tabularx}
\usepackage{longtable}
\usepackage{array}
\usepackage{float}
\usepackage{color}
\usepackage{epstopdf}
\usepackage{mathrsfs}
\usepackage{multirow}
\usepackage[T1]{fontenc}
\usepackage[colorlinks,linkcolor=blue,anchorcolor=blue,citecolor=blue,urlcolor=blue]{hyperref}

\begin{document}

\title{Intrinsic Second-Order Anomalous Hall Effect and Its Application in Compensated
Antiferromagnets}

\author{Huiying Liu}
\affiliation{Research Laboratory for Quantum Materials, Singapore University of Technology and Design, Singapore 487372, Singapore}

\author{Jianzhou Zhao}
\email{jzzhao@swust.edu.cn}
\affiliation{Research Laboratory for Quantum Materials, Singapore University of Technology and Design, Singapore 487372, Singapore}
\affiliation{Co-Innovation Center for New Energetic Materials, Southwest University of Science and Technology, Mianyang 621010, China}

\author{Yue-Xin Huang}
\affiliation{Research Laboratory for Quantum Materials, Singapore University of Technology and Design, Singapore 487372, Singapore}

\author{Weikang Wu}
\address{Research Laboratory for Quantum Materials, Singapore University of Technology and Design, Singapore 487372, Singapore}
\address{Division of Physics and Applied Physics, School of Physical and Mathematical Sciences,
Nanyang Technological University, Singapore 637371, Singapore}

\author{Xian-Lei Sheng}
\affiliation{School of Physics, and Key Laboratory of Micro-nano Measurement-Manipulation and Physics, Beihang University, Beijing 100191, China}

\author{Cong Xiao}
\email{congxiao@hku.hk}
\address{Department of Physics, The University of Hong Kong, Hong Kong, China}
\address{HKU-UCAS Joint Institute of Theoretical and Computational Physics at Hong Kong, China}
\address{Department of Physics, The University of Texas at Austin, Austin, Texas 78712, USA}

\author{Shengyuan A. Yang}
\affiliation{Research Laboratory for Quantum Materials, Singapore University of Technology and Design, Singapore 487372, Singapore}

\begin{abstract}
Response properties that are purely intrinsic to physical systems are of paramount importance in physics research, as they probe
fundamental properties of band structures and allow quantitative calculation and comparison with experiment. For anomalous Hall transport in magnets, an intrinsic effect can appear at the second order to the applied electric field. We show that this intrinsic second-order anomalous Hall effect is associated with an intrinsic band geometric property --- the dipole moment of Berry-connection polarizability (BCP) in momentum space. The effect has scaling relation and symmetry constraints that are distinct from the previously studied extrinsic contributions. Particularly, in antiferromagnets with $\mathcal{PT}$ symmetry, the intrinsic effect dominates. Combined with first-principles calculations,
we demonstrate the first quantitative evaluation of the effect in the antiferromagnet Mn$_{2}$Au. We show that the BCP dipole and the resulting intrinsic second-order conductivity are pronounced around band near degeneracies. Importantly, the intrinsic response exhibits sensitive dependence on the N\'{e}el vector orientation with a $2\pi$ periodicity, which offers a new route for electric detection of the magnetic order
in $\mathcal{PT}$-invariant antiferromagnets.

\end{abstract}
\maketitle

Anomalous Hall effect (AHE) is a fundamental transport phenomenon, in which a
transverse charge current is generated in response to a longitudinal electric field without
external magnetic fields. The underlying mechanisms are classified
into {intrinsic} and {extrinsic} ones, depending on whether or not the mechanism is
related to carrier scattering~\cite{sinitsyn2007,nagaosa2010,xiao2010}. In the study of linear AHE, a great success in the past twenty years is the recognition of the
importance of intrinsic contribution and its connection to a band geometric quantity---the Berry curvature~\cite{jungwirth2002,onoda2002}. Recently, the research on AHE has been extended to the nonlinear regime. Sodemann and Fu~\cite{sodemann2015} proposed an \emph{extrinsic} second-order AHE, which involves the dipole of Berry curvature and is linear in the relaxation time. In fact, before Ref.~\cite{sodemann2015}, an \emph{intrinsic} second-order AHE has been predicted by Gao \emph{et al.}~\cite{gao2014}, but received less attention. Particularly, the physical content of this intrinsic effect have not been fully understood, and furthermore, it has not been explored in any concrete material yet.

Meanwhile, in the field of spintronics, a recent focus is to utilize compensated antiferromagnets for device applications, owing to their advantages like robustness to external magnetic perturbations, absence of stray fields, and ultrafast dynamics~\cite{jungwirth2016,baltz2018,smejkal2018}. Especially, the class of $\mathcal{PT}$-symmetric antiferromagnets have been attracting great interest, as they permit a field-like spin-orbit torque to control the N\'{e}el vectors~\cite{zelezny2014}, which has been successfully demonstrated in materials like CuMnAs~\cite{wadley2016,grzybowski2017,wadley2018,godinho2018} and Mn$_2$Au~\cite{bodnar2018,meinert2018,bodnar2019}. However, an outstanding challenge is how to read out the information, i.e., to detect the N\'{e}el vector orientation in these systems. Conventional magnetic measurements fail due to the absence of net magnetization~\cite{baltz2018}; optical microscopy works~\cite{grzybowski2017,saidl2017,sun2019} but is difficult to incorporate for compact devices; and the approach based on anisotropic magnetoresistance (AMR) effect~\cite{wadley2016,baltz2018} suffers from the limited reading speed and cannot distinguish a 180$^\circ$ reversal~\cite{zelezny2018}. Very recently, Shao \emph{et al.}~\cite{shao2020} suggested that for antiferromagnets with broken $\mathcal{PT}$ such as CuMnSb, the extrinsic second-order AHE could be used to detect the N\'{e}el vector. Unfortunately, this cannot apply for $\mathcal{PT}$-symmetric antiferromagnets, since the Berry curvature and hence the effect are suppressed by the
$\mathcal{PT}$ symmetry.

\renewcommand{\arraystretch}{1.42}

\begin{table*}[pbt]
\begin{centering}
\begin{tabular}{p{0.05\linewidth}p{0.032\linewidth}<{\centering}p{0.032\linewidth}<{\centering}
p{0.032\linewidth}<{\centering}p{0.032\linewidth}<{\centering}p{0.032\linewidth}<{\centering}
p{0.032\linewidth}<{\centering}p{0.032\linewidth}<{\centering}p{0.032\linewidth}<{\centering}p{0.038\linewidth}<{\centering}
p{0.042\linewidth}<{\centering}p{0.043\linewidth}<{\centering}p{0.043\linewidth}<{\centering}p{0.043\linewidth}<{\centering}
p{0.043\linewidth}<{\centering}p{0.042\linewidth}<{\centering}p{0.042\linewidth}<{\centering}p{0.043\linewidth}<{\centering}
p{0.043\linewidth}<{\centering}p{0.043\linewidth}<{\centering}}
\hline\hline
 &  $\mathcal{P}$ & $C_{n}^{z}$ & $C_{n}^{x}$ & $\sigma_{z}$ & $\sigma_{x}$ & $S_{4,6}^{z}$ & $S_{4}^{x}$ & $S_{6}^{x}$ &
$\ \mathcal{T}$ & $\mathcal{PT}$ & $C_{2}^{z}\mathcal{T}$ & $C_{2}^{x}\mathcal{T}$ & $C_{3,6}^{x}\mathcal{T}$ &
$C_{4}^{x}\mathcal{T}$ & $\sigma_{z}\mathcal{T}$ & $\sigma_{x}\mathcal{T}$ & $S_{4,6}^{z}\mathcal{T}$ & $S_{4}^{x}\mathcal{T}$ & $S_{6}^{x}\mathcal{T}$ \\
\hline
$\chi^\text{int}_{yxx}$   & $\times$ & $\times$ & $\times$ &  $\checkmark$ & $\checkmark$ & $\times$ & $\times$  & $\times$ &
 $\ \times$ & $\checkmark$ & $\checkmark$ & $\checkmark$ & $\times$ & $\times$ &
$\times$ & $\times$ & $\times$ & $\times$ & $\times$\\[0.2ex]
$\chi^\text{int}_{xyy}$  & $\times$ & $\times$ & $\checkmark$   & $\checkmark$ & $\times$ & $\times$ & $\checkmark$ & $\times$ &
 $\ \times$ & $\checkmark$ & $\checkmark$ & $\times$ & $\times$ & $\checkmark$ &
$\times$ & $\checkmark$ & $\times$ & $\checkmark$ & $\checkmark$\\[0.2ex]
$\chi^\text{BCD}_{yxx}$   & \multicolumn{8}{c}{\multirow{2}{*}{Same as $\chi^{\text{int}}$ above}}  &
 $\ \checkmark$ & $\times$ & $\times$ & $\times$  & $\times$ & $\times$ &
 $\checkmark$ & $\checkmark$ & $\times$ & $\times$ & $\times$\\[0.2ex]
$\chi^\text{BCD}_{xyy}$  & \multicolumn{8}{c}{} &
 $\ \checkmark$ & $\times$ & $\times$ & $\checkmark$ & $\checkmark$ & $\checkmark$ &
 $\checkmark$ & $\times$ & $\times$ & $\checkmark$ & $\times$\\[0.2ex]

\hline
\hline
\end{tabular}
\par\end{centering}
\caption{Constraints on the in-plane tensor elements of $\chi^{\text{int}}$ and $\chi^{\text{BCD}}$ from point group symmetries. ``$\checkmark$'' (``$\times$'') means the element is symmetry allowed (forbidden). Here, $C_{3,4,6}^{z}\mathcal{T}$ are not included, as they prohibit all of the listed elements.}%
\label{tab:constraint_2D}%
\end{table*}

In this work, we address the above challenge by showing that the intrinsic second-order AHE offers a powerful tool for electrically detecting
N\'{e}el vectors in $\mathcal{PT}$-symmetric antiferromagnets. We show that the intrinsic effect has a quantum origin connected to the dipole moment of the Berry-connection polarizability (BCP) tensor in momentum space. We clarify the symmetry characters of the effect, and point out its dominant role in $\mathcal{PT}$-symmetric antiferromagnets, where all Berry curvature related first and second order Hall responses are forbidden.
Combining the theory with first-principles calculations, we perform the first quantitative evaluation of the intrinsic second-order AHE in
the paradigmatic $\mathcal{PT}$-symmetric antiferromagnet Mn$_2$Au. The result is found to be sizable and sensitive to the N\'{e}el vector with a $2\pi$ periodicity, indicating a precise way to map out the N\'{e}el vector orientation.


{\color{blue}\textit{Intrinsic second-order AHE and BCP dipole.}} The intrinsic contribution to the second-order AHE is most easily derived within
the extended semiclassical theory, which includes field corrections to the band quantities~\cite{gao2014,gao2015,gao2019,xiao2021therm,xiao2021adia}. In particular, the Berry connection acquires a gauge-invariant correction $\bm{\mathcal{A}}^E$ by the applied electric field $\bm{E}$, with
\begin{equation}
  \mathcal{A}^E_a(\bm k)=G_{ab}(\bm k)E_b,
\end{equation}
where the subscripts $a, b, \cdots$ denote Cartesian coordinates (Einstein summation convention assumed), and $G_{ab}$ is the BCP tensor~\cite{gao2014,liu2021}. For a band with index $n$, BCP can be expressed as (we set $e=\hbar=1$)~\cite{sum}
\begin{equation}
G_{ab}^n(\bm k)=2 \text{Re}\sum_{m\neq n}\frac{\mathcal{A}_{a}^{nm}(\boldsymbol{k})\mathcal{A}_{b}^{mn}(\boldsymbol{k})}%
{\varepsilon_{n}(\boldsymbol{k})-\varepsilon_{m}(\boldsymbol{k})},
\label{eq:BPT}%
\end{equation}
where $\mathcal{A}_{a}^{nm}=\langle u_n|i\partial_a|u_m\rangle$ is the usual interband Berry connection, $|u_n\rangle$ is the
unperturbed eigenstate, $\partial_a\equiv \partial_{k_a}$, and $\varepsilon_{n}$ is the unperturbed band energy.

This generates a field-induced Berry curvature $\boldsymbol{\Omega}^{E}=\nabla_{\boldsymbol{k}%
}\times\bm{\mathcal{A}}^E$, which acts like magnetic field in momentum space and leads to an anomalous velocity term $\sim\bm E\times \bm{\Omega}^{E}$ for electrons. This velocity is transverse to the applied $\bm{E}$ field, of $E^2$ order, and independent of scattering, so it results in the intrinsic second-order AHE
current $\bm j^\text{int}$~\cite{gao2014}. By writing $j^\text{int}_a=\chi_{abc}^\text{int} E_b E_c$, we have
\begin{equation}
\chi_{abc}^\text{int}=\int_\text{BZ}\frac{d\boldsymbol{k}}{(2\pi)^{d}}%
\Lambda_{abc}(\boldsymbol{k}), \label{eq:chi_abc}%
\end{equation}
with%
\begin{equation}\label{Lambda}
\Lambda_{abc}(\boldsymbol{k})=-\sum_{n}(\partial_{a}G_{bc}^n-\partial
_{b}G_{ac}^n)f_0,
\end{equation}
where BZ stands for the Brillouin zone, $d$ is the dimensionality of the system, and $f_0$ is the equilibrium Fermi-Dirac distribution. One observes that the effect is indeed intrinsic, free of scattering effects and involving only intrinsic band geometric quantity, and more precisely, the integrand $\Lambda$ represents a
(anti-symmetrized) combination of the momentum-space dipole moment of BCP over the occupied states. Via an integration by parts, it is also clear that the transport is a Fermi surface property, as it should be.

 As its most important character, the intrinsic conductivity tensor $\chi^\text{int}$ here is completely determined by the band structure, hence can be precisely evaluated from first-principles calculations. This is in contrast to the extrinsic second-order response $\chi^\text{BCD}$ from Berry-curvature dipole in Ref.~\cite{sodemann2015}, which is linear in the scattering time $\tau$. This difference also manifests in their different symmetry properties under time reversal operation: $\chi^\text{int}$ is $\mathcal{T}$ odd, whereas $\chi^\text{BCD}$ is $\mathcal{T}$ even. Thus, the intrinsic contribution requires broken $\mathcal{T}$, as in magnets, but the extrinsic one does not. Nevertheless, as mentioned, in $\mathcal{PT}$-symmetric antiferromagnets, $\chi^\text{BCD}$ is forbidden, but $\chi^\text{int}$ is allowed. And in cases where both contributions coexist, they can be distinguished in experiment by their different scaling with $\tau$.

{\color{blue}\textit{Symmetry property.}} We have seen that the intrinsic second-order conductivity $\chi^\text{int}$ is a
$\mathcal{T}$-odd rank-3 tensor. From Eq.~(\ref{Lambda}), it is clear that $\chi^\text{int}_{abc}$ is antisymmetric in its first two indices, which ensures that $j_a^\text{int}E_a=0$, i.e., $\bm j^\text{int}$ is indeed a Hall current.

For most transport experiments, the setup has a planar geometry, with the applied $\bm{E}$ field and the generated current both within the plane (denoted as the $xy$ plane). Then the effect is specified by only two tensor elements, $\chi^\text{int}_{xyy}$ and $\chi^\text{int}_{yxx}$. For $\bm{E}$ field making an angle $\theta$ from the $x$ direction (usually taken to be certain crystal direction), i.e., $\bm E=E(\cos\theta,\sin\theta)$, the measured in-plane second-order intrinsic anomalous Hall current can be expressed as \cite{supp}
\begin{equation}
  j_\text{AH}^{(2)}=\chi_\text{AH} E^2,
\end{equation}
with
\begin{equation}
  \chi_\text{AH}=\chi_{yxx}^\text{int}\cos\theta-\chi_{xyy}^\text{int}\sin\theta.
\end{equation}

The form of $\chi^\text{int}$ is also constrained by the point group symmetry of the system. Given its antisymmetry in the first two indices,
to analyze its symmetry, it is convenient to transform it to an equivalent rank-2 pseudo-tensor
\begin{equation}
  \mathcal{X}_{cd}=\epsilon_{abc}\chi_{abd}^{\text{int}}/2,\label{pseudotensor}
\end{equation}
where $\epsilon_{abc}$ is the Levi-Civita symbol. Then, the constraints from point group symmetries on $\mathcal{X}$ can be derived from
\begin{equation}
\mathcal{X}=\eta_T \det(O)O\mathcal{X} O^{-1},\label{symmetry}%
\end{equation}
where $O$ is a point group operation, and the factor $\eta_T=\pm $ is again associated with the character of $\chi^\text{int}$ being $\mathcal{T}$ odd: $\eta_T=-1$ for primed operations, i.e., the magnetic symmetry operations of the form $R\mathcal{T}$ with $R$ a spatial operation; and $\eta_T=+1$ for non-primed operations. Here, the presence/absence of $\eta_T=-1$ for primed operations is the key distinction between the intrinsic $\chi^\text{int}$ and the extrinsic $\chi^\text{BCD}$ contributions. In Table~\ref{tab:constraint_2D}, we list and compare the constraints of common point group operations on the in-plane $\chi$ tensor elements. One finds that several primed operations, such as $\mathcal{PT}$, $C_2^z \mathcal{T}$, and $S_6^x \mathcal{T}$, can completely suppress $\chi^\text{BCD}$ but allow non-vanishing $\chi^\text{int}$.

\begin{figure}[ptb]
\centering
\includegraphics[width=1\columnwidth]{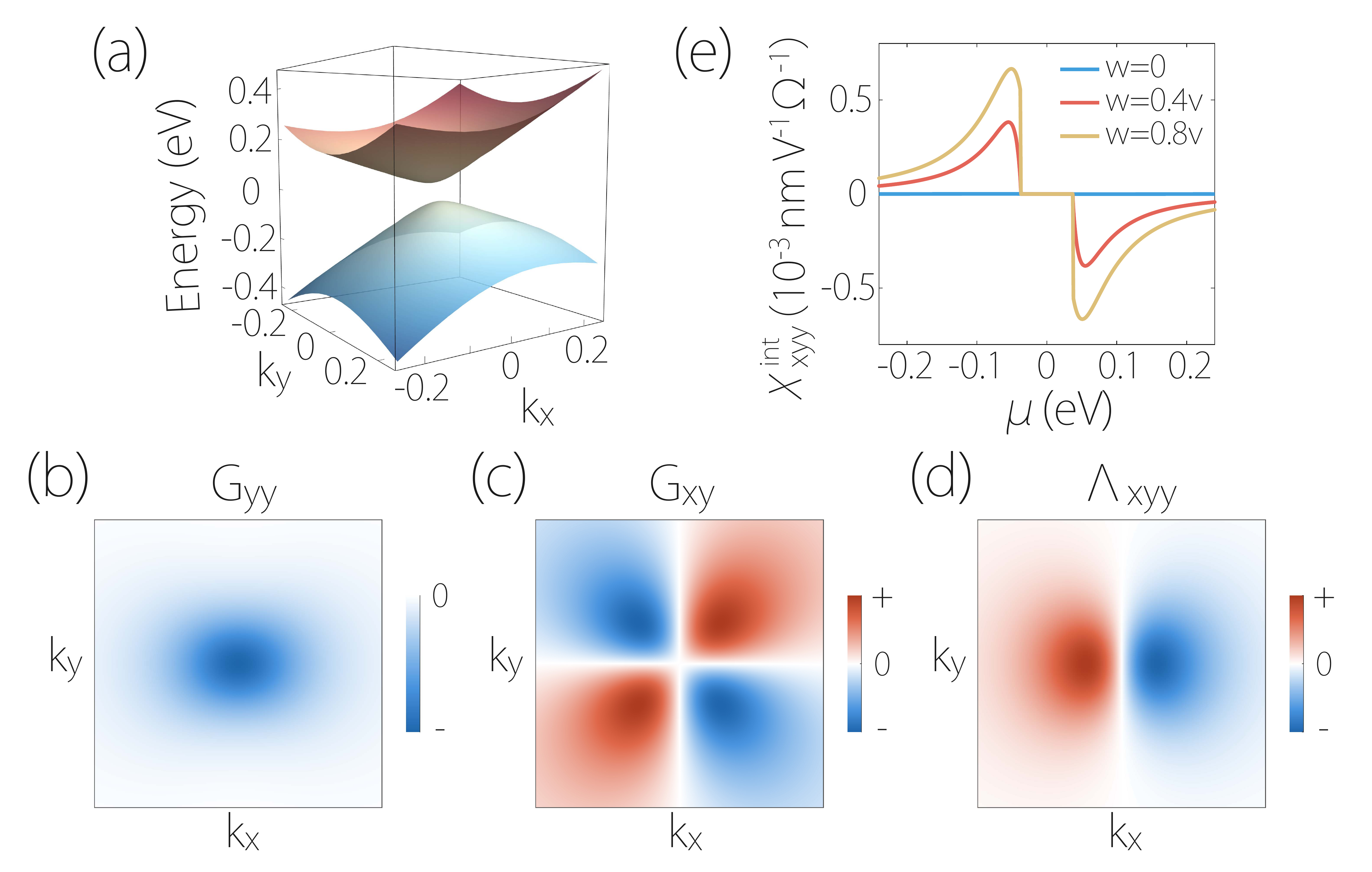}\caption{(a) Band structure of
the 2D Dirac model. (b-d) Distribution of BCP elements (b) $G_{yy}$, (c) $G_{xy}$, and (d) BCP dipole $\Lambda_{xyy}$ in the momentum space for the valence band of the model.
(e) Calculated intrinsic second-order anomalous Hall conductivity $\chi_{xyy}^{\text{int}}$ versus the Fermi energy $\mu$.
In the calculation, we take $v_x\!=\!v_y\!=\!1\times10^{6}$ m/s,  $w=0.4v_x$, and $\Delta=40$ meV.}%
\label{fig:kp}%
\end{figure}

{\color{blue}\textit{2D Dirac model.}} To better understand the features of BCP dipole and the intrinsic second-order AHE, we first apply the theory to study the 2D Dirac model, which is a minimal model for describing a (anti)crossing between two bands.
The model reads
\begin{equation}
H(\boldsymbol{k})=wk_{x}+v_{x}k_{x}\sigma_{x}+v_{y}k_{y}\sigma_{y}+\Delta\sigma_{z},\label{Dirac}
\end{equation}
where $\sigma_i$'s
are the Pauli matrices, $v_i$ are Fermi velocities,
$2\Delta$ is the local gap at the band near degeneracy, the first term represents an energy tilt, and we assume the model parameters are positive and $w/v_x<1$.
The model spectrum is plotted in Fig.~\ref{fig:kp}(a). As the model itself has a $C_2^y \mathcal{T}$ symmetry, the element $\chi_{yxx}^\text{int}$ vanishes,
and we focus on $\chi_{xyy}^\text{int}$. The relevant BCP elements $G_{yy}$ and $G_{xy}$ as well as the BCP dipole $\Lambda_{xyy}$ for the valence band are plotted in Fig.~\ref{fig:kp}. One observes that $G_{yy}$ and $G_{xy}$ respectively show a monopole and a quadrupole pattern, and the resulting $\Lambda_{xyy}$ exhibits a dipole pattern along $k_x$. All these quantities are concentrated around the small-gap region.

The intrinsic second-order anomalous Hall conductivity for this model can be derived as~\cite{supp}
\begin{equation}
\chi_{xyy}^\text{int}=-\frac{ v_{y}\lambda\mu [\mu^{2}+(\lambda^{2}-1)\Delta^{2}]}{8\pi (\mu^{2}+\lambda^{2}\Delta^{2})^{{5}/{2}}}\Theta (|\mu|-\Delta),
\end{equation}
where $\lambda=w/v_{x}$, $\mu$ is the Fermi energy, and $\Theta$ is the step function. In Fig.~\ref{fig:kp}(e), we plot $\chi_{xyy}^\text{int}$ as a function of $\mu$. One can see that the response is pronounced when $\mu$ is close to the small gap region, as band near degeneracies are the main source for generating sizable BCP dipoles. In addition, for this Dirac model, the tilt term plays an important role, as it lowers the symmetry to allow non-vanishing BCP dipole and $\chi^\text{int}$.  Without this term, $\chi^\text{int}$ would vanish identically.

{\color{blue}\textit{Application to Mn$_2$Au.}} As its unique advantage, the intrinsic second-order AHE only depends on the
band structure, thus it can be evaluated in first-principles calculations to yield quantitative predictions for concrete materials.
Here, we consider Mn$_2$Au, which is a paradigmatic example of $\mathcal{PT}$-symmetric antiferromagnets and is under active research in recent years~\cite{wu2012,barthem2013,jourdan2015,bodnar2018,meinert2018,bodnar2019}.

\begin{figure}[ptb]
\centering
\includegraphics[width=1\columnwidth]{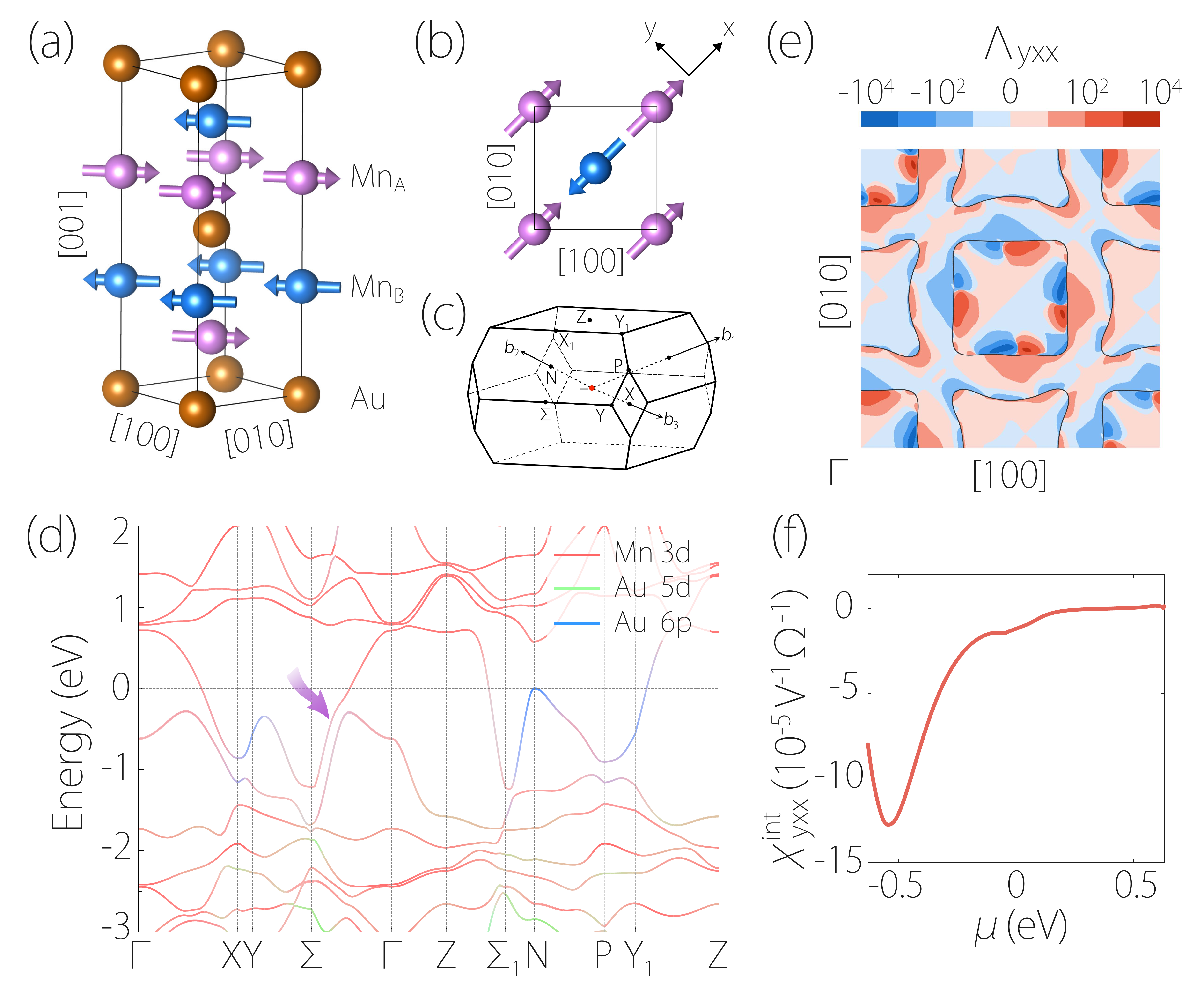}\caption{(a) Crystal structure of Mn$_{2}$Au.
Mn atoms with opposite magnetic moments are marked with two different colors.
Black lines indicate the conventional unit cell. (b) In the ground state, the N\'{e}el vector is along the [110] direction, which is labeled as the $x$ axis here. (c) shows the Brillouin zone. (d) Calculated band structure
of antiferromagnetic Mn$_{2}$Au. The  arrow indicates the small gap region that contributes to the peak in (f).
 (e) Distribution of $\Lambda_{yxx}$  in the $k_{z}=0$ plane of the BZ.
Black lines depict the Fermi surface. (f) Calculated $\chi_{yxx}^{\text{int}}$ versus the Fermi energy $\mu$.  }%
\label{fig:Mn2Au_band}%
\end{figure}

The lattice structure of Mn$_2$Au is shown in Fig.~\ref{fig:Mn2Au_band}, which is tetragonal and belongs to the space group $I4/mmm$ (No.~139).
Experiment shows that Mn$_2$Au is a good metal, with compensated collinear antiferromagnetism and a high N\'{e}el temperature $\!>\!1000$ K~\cite{barthem2013}.
The ground state magnetic configuration is illustrated in Fig~\ref{fig:Mn2Au_band}. The magnetic moments are coupled ferromagnetically within each Mn sheet normal to $c$, whereas two neighboring sheets are antiferromagnetically coupled. The N\'{e}el vector $\bm N$ shows a strong in-plane anisotropy and it prefers the $\langle110\rangle$ direction. Our first-principles calculations based on the density functional theory (DFT) confirm these features (calculation details are presented in \cite{supp}).

The magnetic configuration of Mn$_2$Au belongs to the $Fm'mm$ magnetic space group. Importantly, it preserves $\mathcal{PT}$, which suppresses the
extrinsic contribution $\chi^\text{BCD}$ to the second-order response. In addition, for $\bm N$ along the $[110]$ direction which is chosen to be the $x$ direction here [see Fig.~\ref{fig:Mn2Au_band}(b)], the preserved symmetries $M_x$ and $M_y\mathcal{T}$ dictate that $\chi_{xyy}^\text{int}$ vanishes, and only $\chi_{yxx}^\text{int}$ is needed for describing the in-plane intrinsic second-order Hall transport.

Figure~\ref{fig:Mn2Au_band}(d) shows our calculated band structure for Mn$_2$Au along with the projection onto atomic orbitals. One observes that the low-energy states around the Fermi level are mostly contributed by the Mn-$3d$ orbitals and Au-$5p$ orbitals.
In Fig.~\ref{fig:Mn2Au_band}(e), we plot the BCP dipole $\Lambda_{yxx}$ for the $k_z=0$ plane in the BZ. It is an even function with respect to $M_x$. Again, we see that pronounced contributions are from the band near degeneracies close to the Fermi level, as indicated in Fig.~\ref{fig:Mn2Au_band}(d).
$\chi_{yxx}^\text{int}$ is obtained as the integral of the BCP dipole over the whole BZ. In Fig.~\ref{fig:Mn2Au_band}(f), we further show $\chi_{yxx}^\text{int}$ as a function of the Fermi energy. Without doping, $\chi_{yxx}^\text{int}$ is about $-1.2\times10^{-5}$ V$^{-1}\Omega^{-1}$. Taking the longitudinal resistivity $\rho\sim 7$ $\mu\Omega\ \text{cm}$ obtained from experiment~\cite{jourdan2015}, we estimate that for a sample of lateral size $\sim100$ $\mu$m under an
electric field of 1 kV/cm, the induced intrinsic nonlinear Hall
voltage is $\sim$1 $\mu$V, which can be well probed in experiment. Figure~\ref{fig:Mn2Au_band}(f) also shows that the effect would be greatly enhanced when the Fermi energy is shifted towards $-0.5$ eV, because of the band near degeneracy located at that energy [see Fig.~\ref{fig:Mn2Au_band}(d)].


\begin{figure}[ptb]
\centering
\includegraphics[width=1\columnwidth]{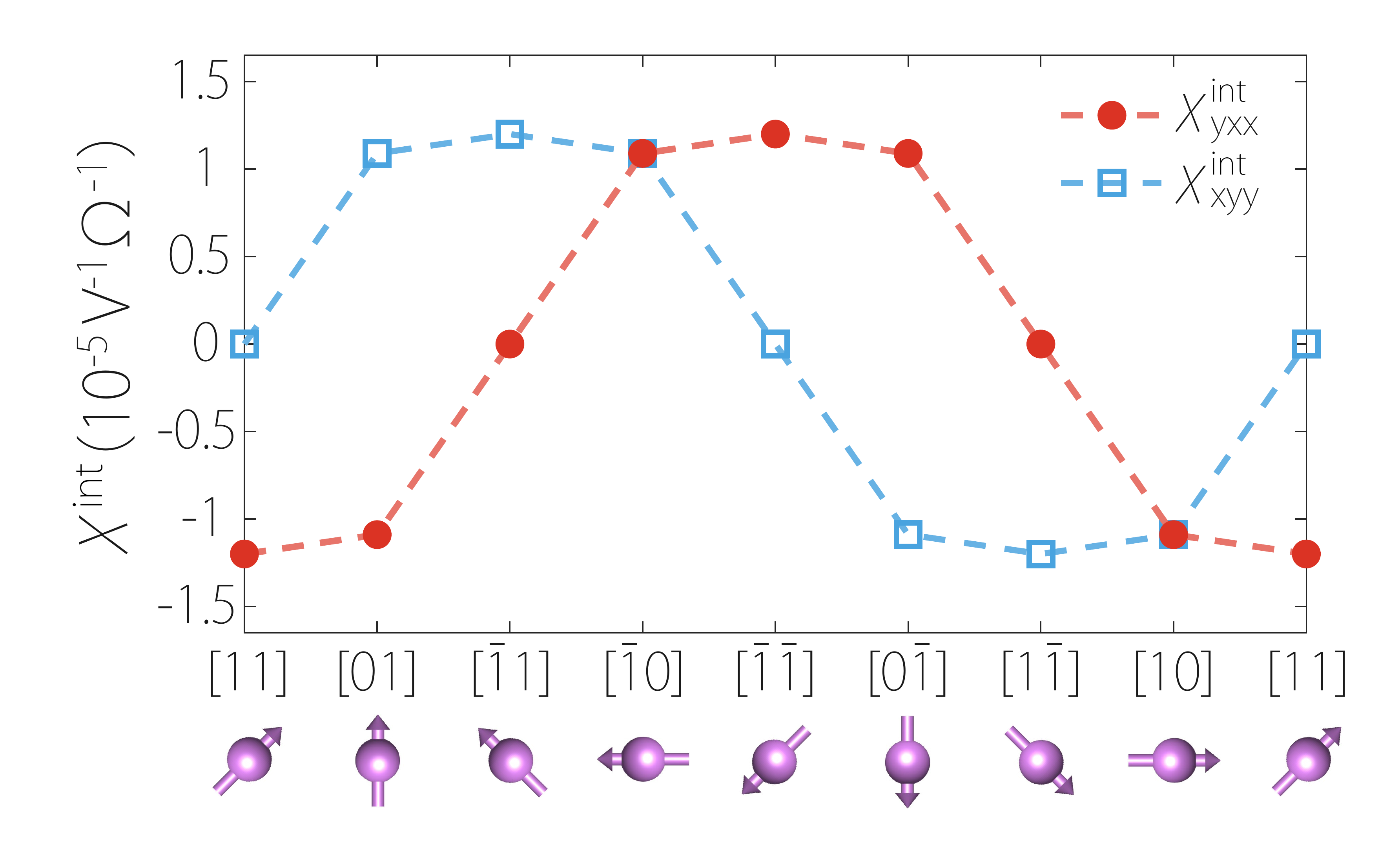}\caption{(a) Calculated intrinsic second-order
conductivity $\chi_{yxx}^{\text{int}}$ and $\chi_{xyy}^{\text{int}}$ of Mn$_{2}$Au when the N\'{e}el vector (denoted by the moment orientation of Mn$_{\text{A}}$)
rotates in the $xy$ plane.
}%
\label{fig:Mn2Au_spin}
\end{figure}

%

Most importantly, we show that the intrinsic second-order AHE sensitively depends on the Neel vector direction, thereby it serves as a powerful tool for detecting $\bm N$. For example, we fix the driving field and the measurement directions to be along $x$ and $y$, respectively. Then the response is specified by $\chi_{yxx}^\text{int}$. Figure~\ref{fig:Mn2Au_spin} shows the variation of $\chi_{yxx}^\text{int}$ when the N\'{e}el vector rotates in the $xy$ plane (here $\bm N$ is defined to be along the moments of the purple-colored sublattice). Importantly, $\chi_{yxx}^\text{int}$ exhibits a $2\pi$ periodicity, meaning that the measurement is capable to fully map out the N\'{e}el vector orientation. This is in contrast to measurement based on linear AMR~\cite{wadley2016,bodnar2018}, which has a $\pi$ periodicity and cannot distinguish a $180^\circ$ reversal. Here, the $180^\circ$ reversal would flip the sign of the signal, as $
\chi^\text{int}(\alpha)=-\chi^\text{int}(\alpha+\pi)$ where $\alpha$ is the polar angle of $\bm N$. In Fig.~\ref{fig:Mn2Au_spin}, we also include the curve for $\chi_{xyy}^\text{int}$. In fact, due to the $C^z_4$ symmetry, $\chi_{xyy}^\text{int}$ is not independent, but related to $\chi_{yxx}^\text{int}$ via
$\chi_{xyy}^\text{int}(\alpha)=-\chi_{yxx}^\text{int}(\alpha-\pi/2)$. Thus, the in-plane intrinsic second-order AHE here can be completely specified by a single tensor element.

{\color{blue}\textit{Discussion.}} We have shown that the intrinsic second-order AHE offers a new route for
probing the BCP dipole, which is an intriguing band geometric quantity, and for detecting N\'{e}el vectors, which is a challenge
in antiferromagnetic spintronics. We demonstrate the first-principles evaluation of the effect for a concrete material. The study can be naturally extended to other materials such as MgMnGe, MnPd$_2$ and  CuMnAs, also including magnets without $\mathcal{PT}$, as long as the effect is symmetry allowed.

We focus on the intrinsic effect in this work. Similar to the linear AHE, there are other extrinsic contributions in the second-order response, but they typically exhibit different behaviors. As mentioned, the Berry curve dipole induced one (and also the conventional Drude contribution) can be distinguished from the intrinsic one by their different scalings with $\tau$~\cite{kang2019,lai2021}. It has also been shown that the extrinsic contributions in the zeroth-order of scattering time from the so-called
coordinate-shift and skew scattering mechanisms are suppressed by the
$\mathcal{PT}$ symmetry~\cite{watanabe2020}.

In practice, the effect can be measured with the standard Hall bar setup as in Refs.~\cite{ma2019,kang2019,lai2021} For $\mathcal{PT}$-symmetric antiferromagnets, it has been experimentally demonstrated that the N\'{e}el vector can be rotated by current pulse via the field-like spin-orbit torques~\cite{wadley2016,grzybowski2017,bodnar2018,wadley2018,godinho2018,meinert2018,bodnar2019}. Combined with the detection scheme by the intrinsic second-order AHE proposed here, it is possible to achieve a full-electric ``write-in'' and ``read-off'' device based on antiferromagnetic platforms, which is a central goal of the field.

\emph{Note added.}  After this work was finalized, a complementary and independent study~\cite{wang2021} appeared, with similar theory and calculation done for a different material.



\bibliography{intNAHE_ref}

\begin{thebibliography}{37}
\bibitem [{sum()}]{sum}%
  \BibitemOpen
  \href@noop {} {}\bibinfo {howpublished} {For degenerate bands, the summed $m$
  is out of the degenerate subspace.}\BibitemShut {Stop}%
\end{thebibliography}

\end{document}